\documentclass[10pt,letterpaper]{article}
\usepackage{opex3}
\usepackage{color}
\usepackage{graphicx}
\usepackage{subfig}
\usepackage[utf8]{inputenc}
\usepackage{siunitx}
\begin{document}

\title{Single-mode squeezing in arbitrary spatial modes}

\author{%
    Marion~Semmler$^{1,2}$, 
    Stefan~Berg-Johansen$^{1,2,*}$, 
    Vanessa~Chille$^{1,2}$,
    Christian~Gabriel$^{1,2}$, 
    Peter~Banzer$^{1,2,3}$, 
    Andrea~Aiello$^{1,2}$, 
    Christoph~Marquardt$^{1,2}$
    and Gerd~Leuchs$^{1,2,3}$%
}

\address{%
$^1$Max Planck Institute for the Science of Light, 
Guenther-Scharowsky-Str.~1/Bldg.~24, 
91058~Erlangen, Germany\\
$^2$Institute for Optics, Information and Photonics, 
University~Erlangen-Nuremberg, 
Staudtstr.~7/B2, 
91058~Erlangen, Germany\\
$^3$Department of Physics, 
University of Ottawa, 
25 Templeton, 
Ottawa, Ontario, K1N 6N5 Canada%
}

\email{$^*$stefan.berg-johansen@mpl.mpg.de} 

\begin{abstract}
As the generation of squeezed states of light has become a standard technique
in laboratories, attention is increasingly directed towards adapting the
optical parameters of squeezed beams to the specific requirements of individual
applications. It is known that imaging, metrology, and quantum information may
benefit from using squeezed light with a tailored transverse spatial mode.
However, experiments have so far been limited to generating only a few squeezed
spatial modes within a given setup. Here, we present the generation of
single-mode squeezing in Laguerre-Gauss and Bessel-Gauss modes, as well as an
arbitrary intensity pattern, all from a single setup using a spatial light
modulator (SLM). The degree of squeezing obtained is limited mainly by the
initial squeezing and diffractive losses introduced by the SLM, while no excess
noise from the SLM is detectable at the measured sideband. The experiment
illustrates the single-mode concept in quantum optics and demonstrates the
viability of current SLMs as flexible tools for the spatial reshaping of
squeezed light.
\end{abstract}



\section{Introduction}
Squeezed states of the electromagnetic field 
\cite{%
loudon-quantum-2000,
leuchs-squeezing-1988,
davidovich-sub-poissonian-1996,
dodonov-nonclassical-2002} 
have been the subject of intense theoretical and experimental study over the
past four decades. As a generalisation of Glauber's coherent states
\cite{glauber-quantum-1963} 
to minimum uncertainty states 
\cite{stoler-equivalence-1970,lu-new-1971},
their experimental realisation 
\cite{slusher-observation-1985}
has provided a striking confirmation of the quantum theory of light. 
From an applied point of view, squeezed states allow the limitations imposed by
quantum uncertainties on the accuracy of optical measurements to be overcome.
It was pointed out early on by Caves \cite{caves-quantum-mechanical-1981} that
the sensitivity of gravitational wave interferometers can be enhanced by
squeezing the vacuum state entering the interferometer's unused port. In
recent years, this idea became reality in the GEO~600 interferometer, where
squeezing is currently used to enhance long-term sensitivity by
\SI{2.0}{\decibel} \cite{grote-first-2013}.
In quantum information science, squeezed states are relevant as a resource for
continuous-variable (CV) entanglement
\cite{braunstein-quantum-2005},
as well as CV quantum key distribution protocols
\cite{hillery-quantum-2000,gottesman-secure-2001,jacobsen-elimination-2014}.

Meanwhile, the recognition of light's transverse spatial degrees of freedom
as an information carrier 
\cite{allen-orbital-1992} 
has made spatial modes relevant for optical implementations of 
quantum information protocols 
\cite{%
sasada-transverse-mode-2003,
groblacher-experimental-2006,
armstrong-programmable-2012}. 
The consideration of the effects of quantum noise on the spatial statistics of
photons also led to the field of quantum imaging 
\cite{kolobov-quantum-2007,brida-experimental-2010}, 
and it was found that the displacement and tilt of a laser beam can be measured
more accurately by interfering squeezed light with higher-order spatial modes
\cite{treps-quantum-2003}.
Finally, it was shown that thermal noise from mirror coatings in gravitational
wave interferometers can be reduced by using higher-order Laguerre-Gauss (LG)
modes instead of the fundamental mode, allowing for higher optical powers and
thus an improved signal-to-noise ratio
\cite{granata-higher-order-2010,fulda-experimental-2010}.
Combining this technique with squeezed light within the stringent parameter
regime required by gravitational wave interferometry presents a formidable
challenge, but would allow the phase sensitivity to increase even further.

Here, we present a proof-of-principle experiment to generate amplitude
squeezing in light beams with arbitrary spatial intensity patterns.
As a demonstration of the setup's versatility, we generate squeezed LG and
Bessel-Gauss (BG) beams of different orders, as well as a complex pattern
containing high spatial frequencies. 
All modes are generated without modifications to the setup. The presented
experiment showcases the possibility of generating practically arbitrary
two-dimensional spatial modes that are single-mode squeezed.

Before detailing the experimental setup, we first review the relation between
spatial modes and quantum states of light, and describe some of the existing
approaches for spatial mode squeezing.

\subsection{Quantum states of light and transverse spatial modes}
\label{sec:quantum_spatial}
In order to make more precise the relationship between spatial modes of the
electromagnetic field and photon statistics, we recall the canonical
quantisation of the transverse electromagnetic field, which considers the modes
of a finite volume with either periodic or reflecting boundaries (eventually to
be taken to infinity). 
The electric field operator can be expanded as
\begin{equation}
    \hat{\vec{E}} (\vec{r},t)
    =
    i \sum_k \sqrt{ \frac{\hbar \omega_k}{2} }
    \left( \vec{u}_k(\vec{r},t) \hat{a}_k
        -
        \vec{u}_k^*( \vec{r}, t ) \hat{a}_k^\dagger
        \right),
    \label{eq:quantised-field}
\end{equation}
where $\hat{a}$ and $\hat{a}^\dagger$ are bosonic ladder operators satisfying
$[\hat{a}_i, \hat{a}_j^\dagger]=\delta_{ij}$, and the functions $\vec{u}_k$
denote mutually orthogonal solutions of the Helmholtz equation that describe
transverse oscillation of the transverse electric fields and can be directly
derived from classical Maxwell equations.
However, the generalised definition of an optical mode permits superpositions
of such solutions to be treated as a single-mode excitation as long as the
field possesses first-order coherence
\cite{%
titulaer-density-1966,
deutsch-basis-independent-1991,
treps-quantum-2005,
smith-photon-2007}.
Thus, as long as the condition of first-order coherence is met, a beam
with a complicated spatial structure may be treated as a single-mode
excitation of the field.

When a squeezed beam interacts with a diffractive optical element (such as an
SLM in the present work), the resulting mode pattern can be determined from the
classical theory of diffraction by considering the plane wave spectrum.
Diffraction does not affect the quantum statistics of the mode \textit{per se}.
Rather, we observe a reduction of the degree of squeezing since we are no
longer able to integrate over the full plane wave spectrum with our detector
(i.e., high diffraction orders result in losses)
\cite{smolka-continuous-wave-2012}.
The effect of such losses on the single-mode squeezing can then be found from
the beam splitter relation to be
$\mathrm{Var}_\mathrm{out}
=
\eta \cdot \mathrm{Var}_{\mathrm{in}} + (1-\eta) \cdot \mathrm{Var}_\mathrm{vac}$.
Here, $\eta$ is the efficiency, $\mathrm{Var}_{\mathrm{in}}$ and
$\mathrm{Var}_\mathrm{out}$ represent the variances of the input and output
beam of the squeezed quadrature, and $\mathrm{Var}_\mathrm{vac}$ is the
vacuum variance (shot-noise). Apart from losses induced by imperfect
reflectivity and absorption one also has to take into account possible sources
of additional classical noise. If the diffractive element were to impose an
unwanted temporal modulation at a frequency $f_N$ (for example, due to
electronic flicker noise in the case of a liquid crystal SLM), the quantum
noise of the detected light mode would be masked by excess noise at the optical
sidebands at $f_0 \pm f_N$, rapidly degrading the observable squeezing.

There are various studies of nonclassical beams in a multimode setting
\cite{%
kumar-degenerate-1994,
opatrny-mode-2002,
brambilla-high-sensitivity-2008,
lopez-multimode-2009,
janousek-optical-2009,
corzo-multi-spatial-mode-2011}. 
Here, we concentrate on single-mode squeezing, which is particularly
suited for applications such as quantum-enhanced interferometry.

\subsection{Existing experimental approaches}

Two main approaches to squeezing a single spatial mode have been
demonstrated so far:

\begin{itemize}
    \item[(1)] {\bf Reshaping}, where a squeezed fundamental TEM$_{00}$ mode is
        generated first and subsequently converted into the desired spatial
        mode. Any conversion loss necessarily reduces the squeezing from the
        initial value. This has been achieved with phase plates
        \cite{treps-quantum-2003,treps-nano-displacement-2004,delaubert-quantum-2006},
        where the wavelength and designated mode are fixed, or with
        special-purpose liquid crystal devices
        \cite{gabriel-tools-2012}, which allow more flexibility in the choice
        of wavelength and mode parameters.
        Another approach uses programmable adaptive optics, for which, although
        capable in principle of generating any mode, squeezing has so far
        only been demonstrated in 1D with Hermite-Gauss HG$_{n0}$ modes
        \cite{morizur-spatial-2010}.
    \item[(2)] {\bf Direct squeezing}, where the nonlinear medium is either
        resonant for the desired spatial mode, or transmissive, as in the case
        of a traveling-wave, or single-pass scheme, a squeezed spatial mode can
        be generated directly. Examples include misaligned OPO cavities
        \cite{lassen-generation-2006,lassen-continuous-2009}
        and photonic crystal fibers
        \cite{gabriel-entangling-2011}. 
        The multi-mode squeezing mentioned above can be achieved with this
        approach when the nonlinear medium does not enforce a particular
        spatial mode.
\end{itemize}

\noindent In this work we take the reshaping approach, using an asymmetric
fiber Sagnac interferometer as a squeezing source for TEM$_{00}$ modes
\cite{schmitt-photon-number-1998} and a spatial light modulator for the
subsequent mode conversion.

\section{Experimental setup} 

\subsection{Squeezing}
\begin{figure}[b!]
\centering
\includegraphics[width=12cm]{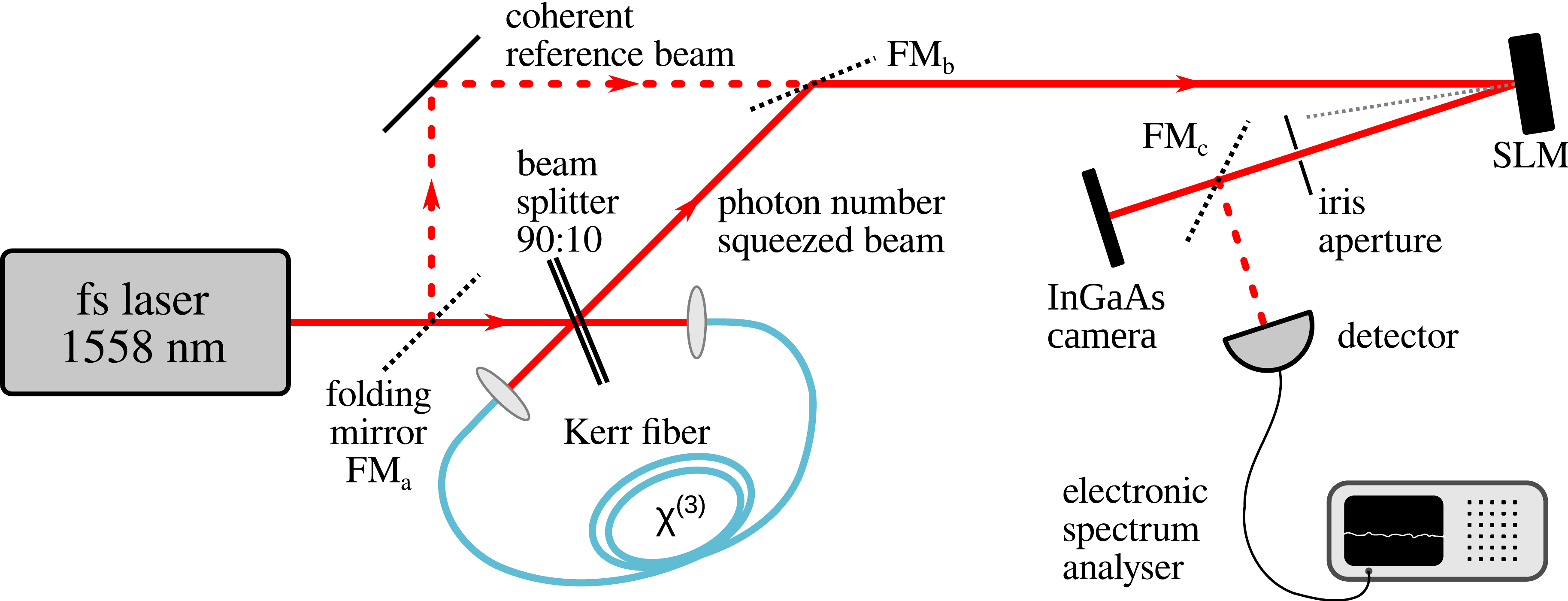}
\caption{%
    \textit{Experimental setup.}
    A femtosecond laser emits pulses of $\SI{220}{\femto\second}$ duration centered
    at $\lambda_0 = \SI{1558}{\nano\metre}$. For squeezed light generation, the
    pulses are split up on a 90:10 beam splitter and launched into a Kerr fiber
    ($\chi^{(3)}$ nonlinearity) of length \SI{3.8}{\metre} in a counter-propagating
    configuration. The exiting pulses typically exhibit
    $\SI{-3.0}{\decibel}$ of amplitude squeezing prior to the SLM. A pair of
    folding mirrors (FM$_\text{a}$, FM$_\text{b}$) allow the squeezer to be
    bypassed to obtain a coherent shot noise reference for squeezing
    measurements. The beam impinges on a reflective SLM. An iris aperture
    selects the 1.~diffraction order (see text for details).  Another folding
    mirror (FM$_\text{c}$) is used to direct the beam either at a InGaAs camera
    for mode inspection or at a detector, whose \SI{9}{\mega\hertz} sideband
    fluctuations and DC amplitude are respectively recorded by an electronic
    spectrum analyser and a volt meter.%
    } \label{fig:sagnac}
\end{figure}

Our light source is a shot-noise limited laser (Origami, Onefive GmbH) emitting
linearly polarised light in $\SI{220}{\femto\second}$ pulses, centered at
a wavelength of $\lambda_0$ = \SI{1558}{\nano\metre}.  Fig.~\ref{fig:sagnac}
shows the asymmetric Sagnac interferometer used to generate amplitude
squeezed light in the initial Gaussian mode. The laser beam is split on an
asymmetric beam splitter with a splitting ratio of 90:10. This results in
a strong and a weak pulse counter-propagating in the polarisation-maintaining
single-mode fiber (FS PM 7811 by 3M). Due to the fiber's nonlinear Kerr effect,
a quadrature squeezing is achieved in the bright pulse that, by means of the
counter-propagating weak pulse, is adjusted to occur in the amplitude
quadrature \cite{schmitt-photon-number-1998}.

\subsection{Mode Conversion}
\label{sec:SLM}
The squeezed Gaussian beam, having a waist of $w_0=\SI{1.32}{\milli\metre}$, is
converted into a higher-order spatial mode by a reflective
liquid-crystal-on-silicon spatial light modulator (LCoS-SLM, Pluto, Holoeye
Photonics AG, 1920x1080~pixels, display optimised for \SI{1550}{\nano\metre},
no anti-reflection coating).
This SLM is designed for phase-only modulation and does not directly
modulate the amplitude. 
The local refractive index is modulated due to the preferential alignment of
the rod-shaped LC molecules with the electric field at each pixel. This affects
only the polarisation component along the long axis of the LC molecules,
leaving the orthogonal polarisation component unmodulated.

We program our SLM with phase patterns consisting
of four contributions:
First, the transverse phase pattern of the theoretical mode function of the
desired mode as described later in this section. Second, a blazed grating
phase which diffracts the modulated beam away from the zeroth order and
transfers the energy mostly into the first diffraction order. This step is
required to spatially separate the modulated light from the approximately 20\%
of incoming light which the SLM effectively does not modulate due to its
limited diffraction efficiency. The grating period of 
$\SI{35}{px} \times \SI{8}{\micro\meter\per px} \approx 180 \, \lambda_0$ is chosen
empirically to maximise diffraction into the first order while enabling sufficient
transverse separation from other diffraction orders in the detection plane at
a distance of \SI{45}{\centi\metre} from the SLM (corresponding to 1/8th of the
Rayleigh length before conversion).
Additionally, a lens phase is added to the hologram. And finally, a binary
circular aperture pattern is multiplied to the entire hologram, restricting 
modulation to the central region only. 

\begin{figure}%
   \centering
   \includegraphics{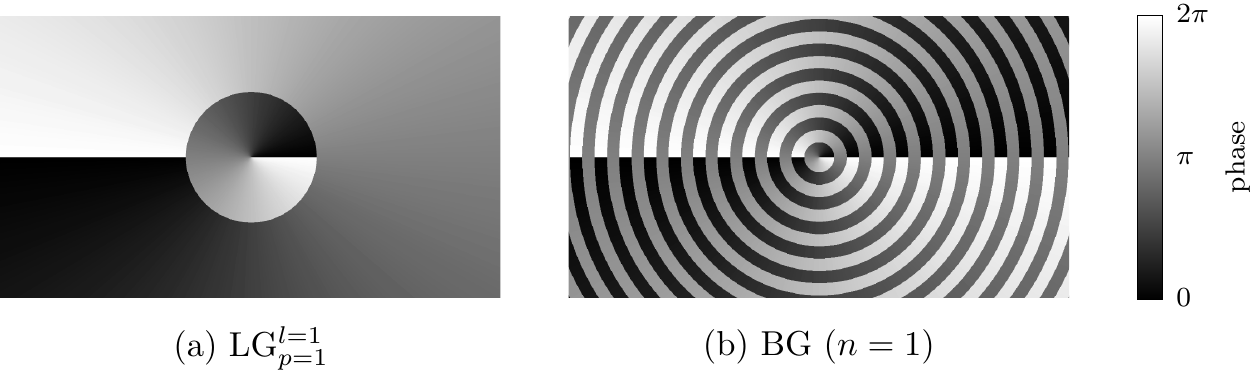}
    \caption{%
        \textit{Example phase patterns.} 
        Basic phase patterns for generating (a) a Laguerre-Gauss beam and (b)
        a Bessel-Gauss beam.  In addition, a blazed grating, kinoform lens and
        aperture are added to each pattern (not shown, see text for details). 
    }
    \label{fig:examples}
\end{figure}

An SLM's important advantage is that it allows for the generation of arbitrary
patterns, i.e.~superpositions of very many basis modes with almost any
combination of coefficients.  To show the versatility of the setup, we generate
amplitude squeezed Laguerre-Gauss beams, Bessel-Gauss beams as well as an
arbitrary pattern.
Laguerre-Gauss (LG) modes are of particular interest as they represent a natural
basis for optical orbital angular momentum \cite{allen-orbital-1999}.
Fig.~\ref{fig:examples}(a) shows the phase pattern required to
generate an LG beam with radial index $p=1$ and helical index $l=1$, where each
photon carries an orbital angular momentum of $\hbar$. 
Bessel beams, too,  exhibit remarkable features: they are non-diffractive and
self-healing \cite{mcgloin-bessel-2005}. These ideal beams extend transversely to
infinity and contain an infinite amount of energy, similar to plane waves.
In a real setting it is hence only possible to generate Bessel-Gauss (BG)
beams, for which the ideal mode function is multiplied by a Gaussian envelope,
while, however, retaining some of its favourable properties.
Fig.~\ref{fig:examples}(b) displays the phase pattern used to generate a BG
beam of order $n=1$.

The two-dimensional spatial Fourier transform of the desired beam is used as
the phase pattern on the SLM. The patterns employed to generate both BG and LG
beams are dominated by their mode functions' defining polynomials, i.e.~the
generalized Laguerre polynomial \cite{allen-orbital-1992} and the $n$th-order
Bessel function of the first kind \cite{mcgloin-bessel-2005}. Every zero in the
radial direction of the polynomial defining the modes results in a phase
discontinuity of $\pi$ of the phase mask (see Fig.~\ref{fig:examples}). The
azimuthal phase consists in a repeated continuous gradient from 0 to 2$\pi$.

\subsection{Measurements}
\label{measurement_variance}
The generated modes are analysed with respect to the quality of the spatial
modes and the quantum noise reduction. The transverse intensity distributions
of the experimentally generated modes are recorded with a Xenics XS-1.7-320
InGaAs camera. As a measure of mode quality, the intensity distribution as
inferred from the theoretical mode function is fitted to a line section of the
measured mode.

The amplitude squeezing is measured by direct photodetection at a sideband
frequency of \SI{9}{\mega\hertz} using an electronic spectrum analyser with
resolution bandwidth \SI{1}{\mega\hertz} and video bandwidth
\SI{3}{\kilo\hertz}. The shot-noise reference level is determined by measuring
the fluctuations of a coherent beam of the same continuous-wave equivalent
optical power in the same spatial mode.  The final squeezing figure is
determined by forming the difference between the respective time-averaged
values. 

For each mode pattern, squeezing is measured in this way both prior to the mode
conversion and in the final converted mode. The mode conversion efficiency
$\eta$ is estimated by comparing the continuous-wave equivalent power of each
diffracted beam behind the SLM to that of the squeezed Gaussian input beam
before the SLM.  The observed reduction in squeezing is compared to the
expected reduction from the variance relation in
Sec.~\ref{sec:quantum_spatial}. Agreement of the two values within their
experimental error indicates that any excess noise power added by the SLM at
the measured sideband lies below the detectable threshold.

\section{Results and Discussion}

\begin{figure}
  \centering
  \includegraphics[width=\textwidth]{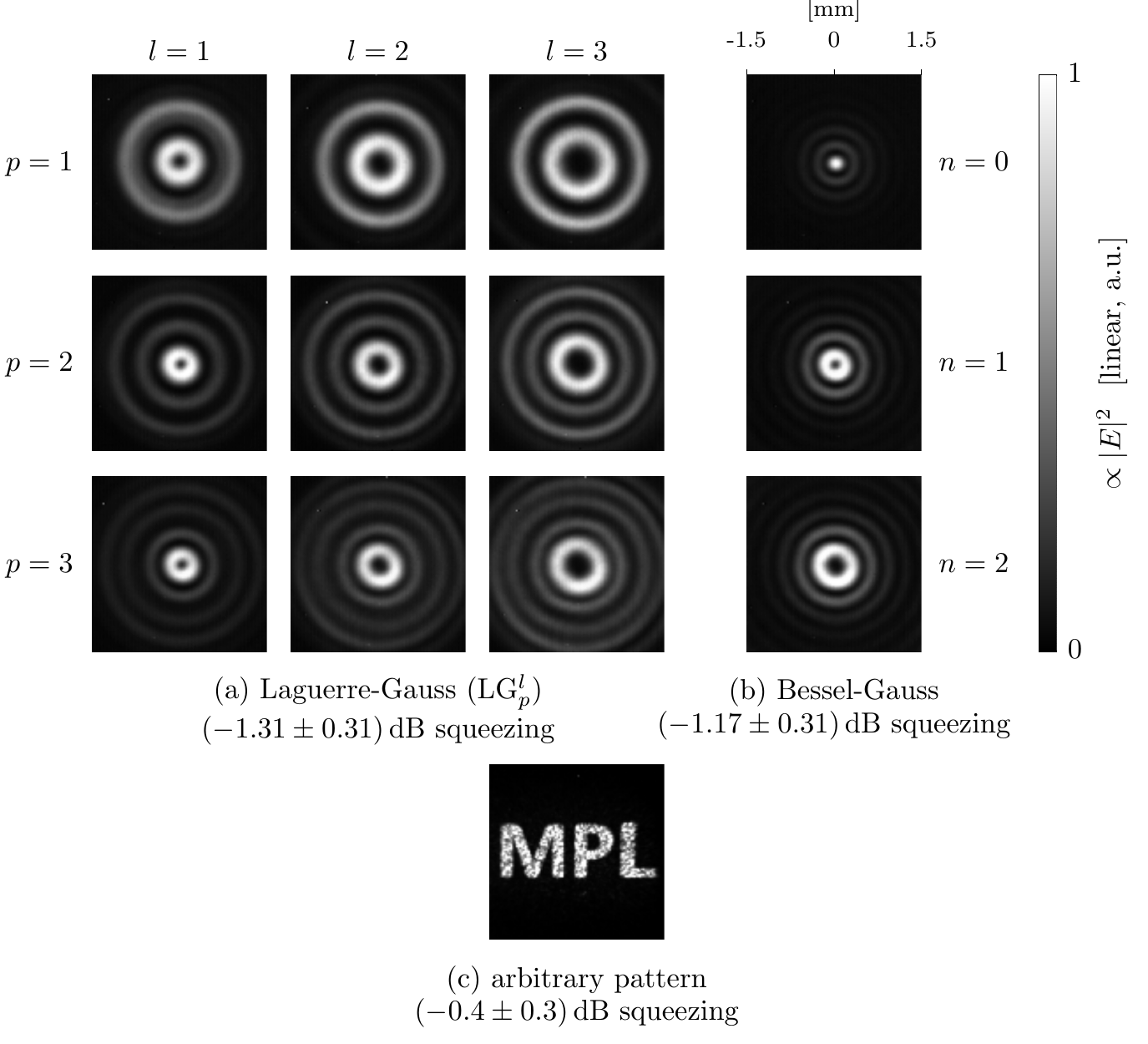}
  \caption{%
    \textit{Squeezed spatial modes.} 
    The average measured amplitude squeezing for each family of modes is shown
    below the panels. Each panel represents a 3$\times$3
    \si{\milli\metre\squared} region in the camera plane. 
    Relative intensity between panels is arbitrary.%
  }
  \label{fig:mode_overview}
\end{figure}

\begin{figure}
    \centering
    \includegraphics[width=\textwidth]{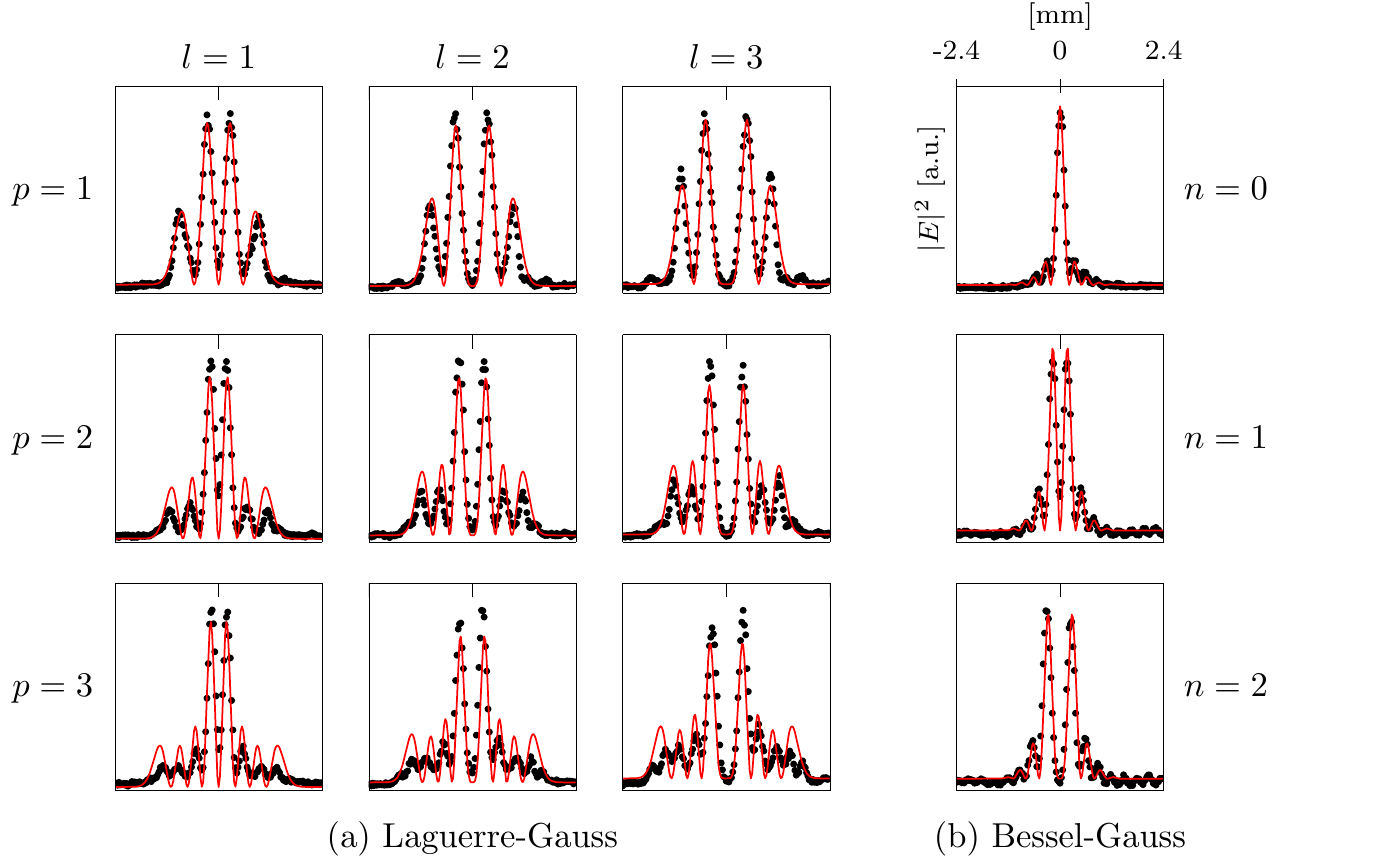}
    \caption{
        \textit{Mode quality.} 
        Measured cross-sectional intensity distributions (dots) with
        fitted theoretical curves (lines) for (a) Laguerre-Gauss beams
        and (b) Bessel-Gauss beams.%
    }
    \label{fig:cross}
\end{figure}

\subsection{Quality of the generated spatial modes}

The observed intensity distributions for LG modes with $p, l \in \{1, 2, 3\}$
and BG modes with $n = 0, 1, 2$ are shown in Fig.~\ref{fig:mode_overview}(a)
and (b), respectively. The intensity distributions are in very good agreement
with the expected beam profiles.
For the same modes, Fig.~\ref{fig:cross}(a) and (b) show the cross-sectional
intensity distributions (black dots) and corresponding fitted curves (red
line).  The fitted curves, derived from the theoretical mode functions,
generally agree well with the observed line sections.  However, for LG modes
with larger values of $p$ the overlap decreases due to the order of
the radial Laguerre polynomial growing with $p$ and becoming increasingly hard
to approximate by phase-only modulation. A method for amplitude and phase
modulation via a single phase-only SLM has been demonstrated by Bolduc {\it et
al.} \cite{bolduc-exact-2013}, based on careful spatial modulation of the
blazed grating depth while compensating the resulting phase aberrations.
However, the technique introduces some additional loss, so that the suitable
trade-off between mode quality and squeezing must be found in accordance with
the requirements of a given application.
More generally, it may be possible to achieve an improvement by measuring and
compensating for any mechanical distortions in the silicon backplane of the
SLM.

\subsection{Optical conversion losses}
We find 
$\eta_d = 0.90(3)$ for the diffraction efficiency and 
$\eta_r = 0.61(2)$ for the reflectivity 
of the SLM. 
The grating efficiency was found to be $\eta_g = 0.91(3)$ in the first order, 
leading to a total efficiency of 
$\eta = \eta_d \eta_r \eta_g = 0.50(3)$ 
with no phase pattern applied.
As shown in tables \ref{tab:lg_results} and \ref{tab:bb_results}, the total
efficiency is reduced further by a few percent for modes with high orders. For the
arbitrary pattern, the efficiency was 0.15.

\subsection{Squeezing in the generated higher-order modes}

\begin{figure}
\centering
\includegraphics[width=\textwidth]{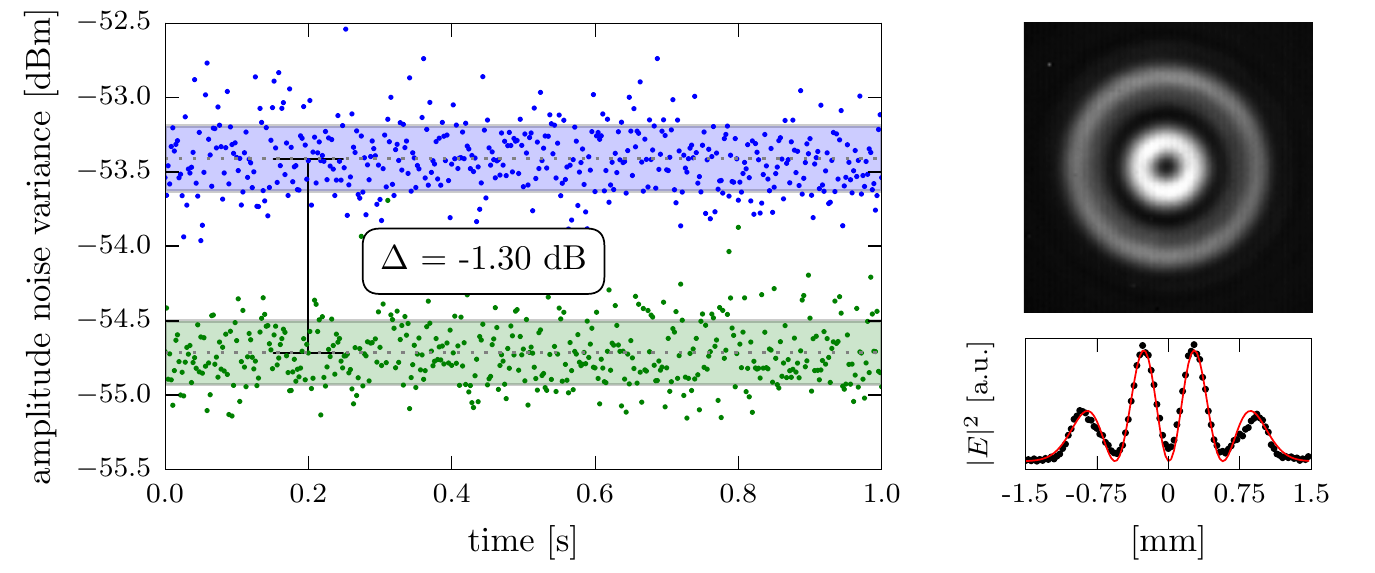}
\caption{%
    \textit{Amplitude squeezing.} 
    Noise trace showing typical amplitude squeezing at \SI{9}{\mega\hertz}
    in the LG$_1^1$ beam. Shaded areas correspond to one standard deviation.%
}
\label{fig:noiseLG11}
\end{figure}

Before the SLM, we typically observe 
$\mathrm{Var}_{\mathrm{in}} = \SI{-3.0(3)}{\decibel}$ of
amplitude squeezing in the fundamental Gauss beam.  The procedure described in
\ref{measurement_variance} to quantify squeezing is demonstrated by example of
an LG$_1^1$ mode in Fig.~\ref{fig:noiseLG11}. Here, a noise reduction of
$\SI{-1.30(30)}{\decibel}$
below the shot-noise level can be seen. Tables \ref{tab:lg_results} and
\ref{tab:bb_results} show a complete list of squeezing values for the LG and BG
beams, respectively. For the arbitrary mode pattern, a noise reduction of
\SI{-0.4(3)}{\decibel} below shot noise was observed.  
Tables~\ref{tab:lg_results}~and~\ref{tab:bb_results} show that for lower-order
modes the efficiencies are much higher ($\approx 50\%$), allowing a typical
squeezing of \SI{-1.3(3)}{\decibel}. All quoted squeezing figures were verified by
attenuation measurements to result from quantum noise reduction. For each
mode, the measured squeezing matches the expected value corresponding to the
measured total efficiency $\eta$ of the SLM for the respective mode to within
the experimental accuracy. In other words, the upper bound for excess noise at
this frequency is lower than the error bars of the measurement. We conclude
that no detectable excess noise was added by the SLM in the mode conversion
process at the \SI{9}{\mega\hertz} sideband.

\begin{table}
\centering
\begin{tabular}{c c c c c}
\hline 
 Radial index	& Helical index & Efficiency & Squeezing \\
$p$	&$l$& $\eta$    & $\Delta$ \\ \hline \hline
1	& 1 & 52.0\%	& \SI{-1.34\pm 0.32}{\decibel} \\
1	& 2 & 51.3\%	& \SI{-1.32\pm 0.31}{\decibel} \\
1	& 3 & 50.1\%	& \SI{-1.34\pm 0.31}{\decibel} \\ \hline
2	& 1 & 51.2\%	& \SI{-1.34\pm 0.30}{\decibel} \\
2	& 2 & 51.2\%	& \SI{-1.31\pm 0.32}{\decibel} \\
2	& 3 & 51.0\%	& \SI{-1.27\pm 0.32}{\decibel} \\ \hline
3	& 1 & 50.3\%	& \SI{-1.30\pm 0.30}{\decibel} \\
3	& 2 & 50.0\%	& \SI{-1.30\pm 0.31}{\decibel} \\
3	& 3 & 50.0\%	& \SI{-1.28\pm 0.31}{\decibel} \\
\hline
\end{tabular}
\caption{Experimental conversion efficiency and measured squeezing in LG$^l_p$ beams.}
\label{tab:lg_results}
\end{table}

\begin{table}
\centering
\begin{tabular}{c c c c}
\hline 
Order & Efficiency & Squeezing \\
$n$ & $\eta$    & $\Delta$ \\ \hline \hline
0 	& 47.0\%	& \SI{-1.17\pm 0.31}{\decibel} \\
1 	& 47.3\%	& \SI{-1.17\pm 0.31}{\decibel} \\
2 	& 47.5\%	& \SI{-1.17\pm 0.31}{\decibel} \\
\hline
\end{tabular}
\caption{Experimental conversion efficiency and measured squeezing in BG beams.}
\label{tab:bb_results}
\end{table}

\section{Summary}
We have shown that a commercially available SLM can be used to transfer
squeezing from the fundamental transverse mode of an optical field into
arbitrary higher-order modes.
With this approach, different spatial modes can be generated simply by
applying a different phase pattern to the SLM with no further modifications
to the setup. In principle, the range of achievable spatial modes is unlimited
(up to the resolution of the SLM), but there is a trade-off between mode
quality and conversion efficiency, which ultimately affects the observable
squeezing in the output mode. In all cases, the observed reduction of squeezing
was consistent with linear losses, ruling out excess noise from the SLM at the
\SI{9}{\mega\hertz} sideband investigated.  Our work provides a direct
illustration of the generalised single-mode concept in quantum optics and shows
that applications requiring squeezed light in tailored spatial modes are within
reach of commercially available technology.


\begin{thebibliography}{10}
\newcommand{\enquote}[1]{``#1''}

\bibitem{loudon-quantum-2000}
R.~Loudon, \enquote{The quantum theory of light,} (Oxford University Press, 2000).

\bibitem{leuchs-squeezing-1988}
G.~Leuchs, \enquote{Squeezing the quantum fluctuations of light,} Contemp.
  Phys. \textbf{29}, 299 (1988).

\bibitem{davidovich-sub-poissonian-1996}
L.~Davidovich, \enquote{Sub-{Poissonian} processes in quantum optics,} Rev.
  Mod. Phys. \textbf{68}, 127 (1996).

\bibitem{dodonov-nonclassical-2002}
V.~V. Dodonov, \enquote{Nonclassical states in quantum optics: a squeezed
  review of the first 75 years,} J. Opt. B: Quantum Semiclass. Opt. \textbf{4}, R1 (2002).

\bibitem{glauber-quantum-1963}
R.~Glauber, \enquote{The {Quantum} {Theory} of {Optical} {Coherence},} Phys.
  Rev. \textbf{130}, 2529 (1963).

\bibitem{stoler-equivalence-1970}
D.~Stoler, \enquote{Equivalence {Classes} of {Minimum} {Uncertainty}
  {Packets},} Phys. Rev. D \textbf{1}, 3217 (1970).

\bibitem{lu-new-1971}
E.~Y.~C. Lu, \enquote{New coherent states of the electromagnetic field,} Lett.
  Nuovo Cimento Series 2 \textbf{2}, 1241 (1971).

\bibitem{slusher-observation-1985}
R.~Slusher, L.~Hollberg, B.~Yurke, J.~Mertz, and J.~Valley,
  \enquote{Observation of {Squeezed} {States} {Generated} by {Four}-{Wave}
  {Mixing} in an {Optical} {Cavity},} Phys. Rev. Lett. \textbf{55}, 2409
  (1985).

\bibitem{caves-quantum-mechanical-1981}
C.~M. Caves, \enquote{Quantum-mechanical noise in an interferometer,} Phys.
  Rev. D \textbf{23}, 1693 (1981).

\bibitem{grote-first-2013}
H.~Grote, K.~Danzmann, K.~L. Dooley, R.~Schnabel, J.~Slutsky, and H.~Vahlbruch,
  \enquote{First {Long}-{Term} {Application} of {Squeezed} {States} of {Light}
  in a {Gravitational}-{Wave} {Observatory},} Phys. Rev. Lett. \textbf{110},
  181101 (2013).

\bibitem{braunstein-quantum-2005}
S.~L. Braunstein and P.~van Loock, \enquote{Quantum information with continuous
  variables,} Rev. Mod. Phys. \textbf{77}, 513 (2005).

\bibitem{hillery-quantum-2000}
M.~Hillery, \enquote{Quantum cryptography with squeezed states,} Phys. Rev. A
  \textbf{61}, 022309 (2000).

\bibitem{gottesman-secure-2001}
D.~Gottesman and J.~Preskill, \enquote{Secure quantum key distribution using
  squeezed states,} Phys. Rev. A \textbf{63}, 022309 (2001).

\bibitem{jacobsen-elimination-2014}
C.~S. Jacobsen, L.~S. Madsen, V.~C. Usenko, R.~Filip, and U.~L. Andersen,
  \enquote{Elimination of information leakage in quantum information channels,}
  arXiv:1408.4566  (2014).

\bibitem{allen-orbital-1992}
L.~Allen, M.~W. Beijersbergen, R.~Spreeuw, and J.~Woerdman, \enquote{Orbital
  angular momentum of light and the transformation of {Laguerre}-{Gaussian}
  laser modes,} Phys. Rev. A \textbf{45}, 8185 (1992).

\bibitem{sasada-transverse-mode-2003}
H.~Sasada and M.~Okamoto, \enquote{Transverse-mode beam splitter of a light
  beam and its application to quantum cryptography,} Phys. Rev. A \textbf{68},
  012323 (2003).

\bibitem{groblacher-experimental-2006}
S.~Gröblacher, T.~Jennewein, A.~Vaziri, G.~Weihs, and A.~Zeilinger,
  \enquote{Experimental quantum cryptography with qutrits,} New J. Phys. \textbf{8}, 75 (2006).

\bibitem{armstrong-programmable-2012}
S.~Armstrong, J.-F. Morizur, J.~Janousek, B.~Hage, N.~Treps, P.~K. Lam, and
  H.-A. Bachor, \enquote{Programmable multimode quantum networks,} Nature
  Commun. \textbf{3}, 1026 (2012).

\bibitem{kolobov-quantum-2007}
M.~Kolobov, \enquote{Quantum imaging,} (Springer, 2007).

\bibitem{brida-experimental-2010}
G.~Brida, M.~Genovese, and I.~Ruo~Berchera, \enquote{Experimental realization
  of sub-shot-noise quantum imaging,} Nature Photon. \textbf{4}, 227 (2010).

\bibitem{treps-quantum-2003}
N.~Treps, N.~Grosse, W.~P. Bowen, C.~Fabre, H.-A. Bachor, and P.~K. Lam,
  \enquote{A quantum laser pointer.} Science \textbf{301}, 940 (2003).

\bibitem{granata-higher-order-2010}
M.~Granata, C.~Buy, R.~Ward, and M.~Barsuglia, \enquote{Higher-{Order}
  {Laguerre}-{Gauss} {Mode} {Generation} and {Interferometry} for
  {Gravitational} {Wave} {Detectors},} Phys. Rev. Lett. \textbf{105}, 231102
  (2010).

\bibitem{fulda-experimental-2010}
P.~Fulda, K.~Kokeyama, S.~Chelkowski, and A.~Freise, \enquote{Experimental
  demonstration of higher-order {Laguerre}-{Gauss} mode interferometry,} Phys.
  Rev. D \textbf{82}, 012002 (2010).

\bibitem{titulaer-density-1966}
U.~M. Titulaer and R.~J. Glauber, \enquote{Density operators for coherent
  fields,} Phys. Rev. \textbf{145}, 1041 (1966).

\bibitem{deutsch-basis-independent-1991}
I.~H. Deutsch, \enquote{A basis-independent approach to quantum optics,} Am. J.
  Phys. \textbf{59}, 834 (1991).

\bibitem{treps-quantum-2005}
N.~Treps, V.~Delaubert, A.~Maître, J.~Courty, and C.~Fabre, \enquote{Quantum
  noise in multipixel image processing,} Phys. Rev. A \textbf{71}, 013820
  (2005).

\bibitem{smith-photon-2007}
B.~J. Smith and M.~G. Raymer, \enquote{Photon wave functions, wave-packet
  quantization of light, and coherence theory,} New J. Phys. \textbf{9}, 414
  (2007).

\bibitem{smolka-continuous-wave-2012}
S.~Smolka, J.~R. Ott, A.~Huck, U.~L. Andersen, and P.~Lodahl,
  \enquote{Continuous-wave spatial quantum correlations of light induced by
  multiple scattering,} Phys. Rev. A \textbf{86}, 033814 (2012).

\bibitem{kumar-degenerate-1994}
P.~Kumar and M.~I. Kolobov, \enquote{Degenerate four-wave mixing as a source
  for spatially-broadband squeezed light,} Opt. Commun. \textbf{104}, 374
  (1994).

\bibitem{opatrny-mode-2002}
T.~Opatrny, N.~Korolkova, and G.~Leuchs, \enquote{Mode structure and photon
  number correlations in squeezed quantum pulses,} Phys. Rev. A \textbf{66},
  053813 (2002).

\bibitem{brambilla-high-sensitivity-2008}
E.~Brambilla, L.~Caspani, O.~Jedrkiewicz, L.~Lugiato, and A.~Gatti,
  \enquote{High-sensitivity imaging with multi-mode twin beams,} Phys. Rev. A
  \textbf{77}, 053807 (2008).

\bibitem{lopez-multimode-2009}
L.~Lopez, B.~Chalopin, A.~de~la Souchère, C.~Fabre, A.~Maître, and N.~Treps,
  \enquote{Multimode quantum properties of a self-imaging optical parametric
  oscillator: {Squeezed} vacuum and {Einstein}-{Podolsky}-{Rosen}-beams
  generation,} Phys. Rev. A \textbf{80}, 043816 (2009).

\bibitem{janousek-optical-2009}
J.~Janousek, K.~Wagner, J.-F. Morizur, N.~Treps, P.~K. Lam, C.~C. Harb, and
  H.-A. Bachor, \enquote{Optical entanglement of co-propagating modes,} Nature
  Photon. \textbf{3}, 399 (2009).

\bibitem{corzo-multi-spatial-mode-2011}
N.~Corzo, A.~M. Marino, K.~M. Jones, and P.~D. Lett,
  \enquote{Multi-spatial-mode single-beam quadrature squeezed states of light
  from four-wave mixing in hot rubidium vapor,} Opt. Express \textbf{19}, 21358
  (2011).

\bibitem{treps-nano-displacement-2004}
N.~Treps, N.~Grosse, W.~P. Bowen, M.~T.~L. Hsu, A.~Maître, C.~Fabre, H.-A.
  Bachor, and P.~K. Lam, \enquote{Nano-displacement measurements using
  spatially multimode squeezed light,} J. Opt. B: Quantum Semiclass. Opt.
  \textbf{6}, 664 (2004).

\bibitem{delaubert-quantum-2006}
V.~Delaubert, N.~Treps, C.~C. Harb, P.~K. Lam, and H.-A. Bachor,
  \enquote{Quantum measurements of spatial conjugate variables: displacement
  and tilt of a {Gaussian} beam.} Opt. Lett. \textbf{31}, 1537 (2006).

\bibitem{gabriel-tools-2012}
C.~Gabriel, A.~Aiello, S.~Berg-Johansen, C.~Marquardt, and G.~Leuchs,
  \enquote{Tools for detecting entanglement between different degrees of
  freedom in quadrature squeezed cylindrically polarized modes,} Eur. Phys. J.
  D \textbf{66}, 172 (2012).

\bibitem{morizur-spatial-2010}
J.-F. Morizur, S.~Armstrong, N.~Treps, J.~Janousek, and H.-A. Bachor,
  \enquote{Spatial reshaping of a squeezed state of light,} Eur. Phys. J. D
  \textbf{61}, 237 (2010).

\bibitem{lassen-generation-2006}
M.~Lassen, V.~Delaubert, C.~C. Harb, P.~K. Lam, N.~Treps, and H.-A. Bachor,
  \enquote{Generation of {Squeezing} in {Higher} {Order} {Hermite}-{Gaussian}
  {Modes} with an {Optical} {Parametric} {Amplifier},} J. Eur. Opt. Soc.-Rapid
  \textbf{1}, 06003 (2006).

\bibitem{lassen-continuous-2009}
M.~Lassen, G.~Leuchs, and U.~L. Andersen, \enquote{Continuous {Variable}
  {Entanglement} and {Squeezing} of {Orbital} {Angular} {Momentum} {States},}
  Phys. Rev. Lett. \textbf{102}, 163602 (2009).

\bibitem{gabriel-entangling-2011}
C.~Gabriel, A.~Aiello, W.~Zhong, T.~Euser, N.~Joly, P.~Banzer, M.~Förtsch,
  D.~Elser, U.~L. Andersen, C.~Marquardt, P.~S. Russell, and G.~Leuchs,
  \enquote{Entangling {Different} {Degrees} of {Freedom} by {Quadrature}
  {Squeezing} {Cylindrically} {Polarized} {Modes},} Phys. Rev. Lett.
  \textbf{106}, 060502 (2011).

\bibitem{schmitt-photon-number-1998}
S.~Schmitt, J.~Ficker, M.~Wolff, F.~König, A.~Sizmann, and G.~Leuchs,
  \enquote{Photon-{Number} {Squeezed} {Solitons} from an {Asymmetric}
  {Fiber}-{Optic} {Sagnac} {Interferometer},} Phys. Rev. Lett. \textbf{81},
  2446 (1998).

\bibitem{allen-orbital-1999}
L.~Allen, M.~Padgett, and M.~Babiker, \enquote{{The} {Orbital} {Angular}
  {Momentum} of {Light},} Prog. Opt. \textbf{39}, 291 (1999).

\bibitem{mcgloin-bessel-2005}
D.~McGloin and K.~Dholakia, \enquote{Bessel beams: {Diffraction} in a new
  light,} Contemp. Phys. \textbf{46}, 15 (2005).

\bibitem{bolduc-exact-2013}
E.~Bolduc, N.~Bent, E.~Santamato, E.~Karimi, and R.~W. Boyd, \enquote{Exact
  solution to simultaneous intensity and phase encryption with a single
  phase-only hologram,} Opt. Lett. \textbf{38}, 3546 (2013).

\end{thebibliography}
\end{document}